\documentclass[12pt,fleqn]{article}
\usepackage{epsfig}
\usepackage[dvips]{color}
\begin{document}
\title{Electroweak phase transitions in the MSSM with an extra $U(1)'$}
\author{S.W. Ham$^{(1)}$, E.J. Yoo$^{(2)}$, and S.K. Oh$^{(1,2)}$
\\
\\
{\it $^{(1)}$ Center for High Energy Physics, Kyungpook National University, }
\\
{\it Daegu 702-701, Korea}
\\
{\it $^{(2)}$ Department of Physics, Konkuk University, Seoul 143-701, Korea}
\\
\\
}
\date{}
\maketitle
\begin{abstract}
We investigate the possibility of electroweak phase transition
in the minimal supersymmetric standard model (MSSM) with an extra $U(1)'$.
This model has two Higgs doublets and a singlet,
in addition to a singlet exotic quark superfield.
We find that at the one-loop level this model may accommodate the electroweak phase transitions
that are strongly first-order in a reasonably large region of the parameter space.
In the parameter region where the phase transitions take place,
we observe that the lightest scalar Higgs boson has a smaller mass
when the strength of the phase transition becomes weaker.
Also, the other three heavier neutral Higgs bosons get more large masses
when the strength of the phase transition becomes weaker.
\end{abstract}
\vfil\eject

\section{INTRODUCTION}

The baryon asymmetry of the universe can be dynamically generated during the evolution of the universe, if the mechanism of baryogenesis satisfies the three Sakharov conditions [1].
The three Sakharov conditions are: the presence of baryon number violation, the violation of both C and CP, and a deviation from thermal equilibrium.
It is known that the universe can escape out of the thermal equilibrium by means of electroweak phase transition, which should be strongly first-order in order to ensure sufficient deviation from thermal equilibrium to generate the baryon asymmetry that is observed today.
However, it has been already recognized that the Standard Model (SM) has some difficulty to realize the desired electroweak phase transition.
The present experimental lower bound on the mass of the SM Higgs boson does not allow the electroweak phase transition to be strongly first-order [2, 3].
The electroweak phase transition is weakly first-order or higher order in the SM.
Thus, the SM is inadequate to generate sufficient baryon asymmetry.
Moreover, the amount CP violation in the Cabibbo-Kobayashi-Maskawa (CKM) matrix
is too small to account for the baryon asymmetry of the observed universe [4].

Consequently, new physical models beyond the SM have extensively been studied for the possibility of reasonable explanation of the baryon asymmetry of the universe.
Especially, the low energy supersymmetric models have been studied widely within the context of electroweak baryogenesis [5-7].
The simplest supersymmetric model that includes the SM is the minimal supersymmetric standard model (MSSM), which possesses in its superpotential the $\mu$ term that accounts for the mixing between two Higgs doublets.
The $\mu$ parameter, which has the mass dimension, causes some problem with respect to its energy scale [8].
Several possibilities have been investigated in the literature to solve the so-called $\mu$ problem [9-12].
Introducing an additional $U(1)'$ to the MSSM is one of the plausible explanations for the $\mu$ problem of the MSSM.

The MSSM with an extra $U(1)'$ can not only solve the $\mu$ problem but we will show that it can also overcome the difficulties that the SM encounters when the SM tries to satisfy the Sakharov conditions.
This model can accommodate sufficient CP violation, because it possesses other sources of CP violation besides the CKM matrix.
It is possible to realize the explicit CP violation in this model by means of complex CP phases arising from the soft SUSY breaking terms [12].

Then, it is the purpose of this paper to show that this model indeed allows the strongly first-order electroweak phase transitions such that it can successfully explain the baryogenesis.
The characteristics of the electroweak phase transitions are determined essentially by the temperature-dependent part of the Higgs potential.
We construct the full temperature-dependent Higgs potential at the one-loop level, and examine if the electroweak phase transition may be strongly first-order.
Two methods are employed for the construction of the temperature-dependent Higgs potential.
One method assumes that the critical temperature at which the electroweak phase transition occurs is relatively high, thus the temperature-dependent effective potential is approximated by retaining only terms proportional to $T^2$, whereas the other method carries out numerically exact integrations of the temperature-dependent effective potential.
The thermal effects of particles whose masses are comparatively smaller than the critical temperature are included at the one-loop level in the former method, whereas the particle content is different in the latter method.

Either way, we obtain almost the same physical results.
Unlike the MSSM, this model allows a strongly first-order electroweak phase transition in a wide region of the parameter space, and the first-order electroweak phase transition can be strong enough without requiring a light stop quark.
An interesting behavior of this model with respect to the strongly first-order electroweak phase transition is that the mass of the lightest neutral Higgs boson becomes larger when the phase transition gets stronger.
On the other hand, the masses of the other three neutral Higgs bosons become smaller when the phase transition gets stronger.

\section{ZERO TEMPERATURE}

The MSSM with an extra $U(1)'$ accommodates in its Higgs sector two Higgs doublets $H_1 = (H_1^0, H_1^-)$, $H_2=(H_2^+, H_2^0)$, and one Higgs singlet, $S$.
In terms of these Higgs fields, the relevant part of the superpotential of this model may be written as
\begin{equation}
    W \approx  h_t Q H_2 t_R^c + h_b Q H_1 b_R^c + h_k S D_L {\bar D}_R - \lambda S H_1^T \epsilon H_2  \ ,
\end{equation}
where we take into account only the third generation:
$t_R^c$ and $b_R^c$ are, respectively, the right-handed singlet top and bottom quark superfields,
$D_R$ is the right-handed singlet exotic quark (a vector-like down quark) superfield,
$Q$ is the left-handed $SU(2)$ doublet quark superfield of the third generation,
and $D_L$ is the left-handed singlet exotic quark superfield.
Further, $h_t$, $h_b$ and $h_k$ are, respectively, the dimensionless Yukawa coupling coefficients of top, bottom, and exotic quark  superfields, and $\epsilon$ is an antisymmetric $2 \times 2$  matrix with $\epsilon_{12} = 1$.

From the superpotential, at zero temperature, we can construct the Higgs potential at the tree level, which may be read as
\begin{equation}
    V_0 = V_F + V_D + V_{\rm S} \ ,
\end{equation}
where
\begin{eqnarray}
    V_F & = & |\lambda|^2 [(|H_1|^2 + |H_2|^2) |S|^2 + |H_1^T \epsilon  H_2|^2]  \ , \cr
    V_D & = & {g_2^2 \over 8} (H_1^{\dagger} \vec\sigma H_1 + H_2^{\dagger} \vec\sigma H_2)^2
        + {g_1^2 \over 8} (|H_1|^2 - |H_2|^2)^2 \cr
        & &\mbox{}+ {g^{'2}_1 \over 2} ( {\tilde Q}_1 |H_1|^2 + {\tilde Q}_2 |H_2|^2
        + {\tilde Q}_3 |S|^2)^2 \ , \cr
    V_{\rm S} & = & m_1^2 |H_1|^2 + m_2^2 |H_2|^2 + m_3^2 |S|^2
        - [\lambda A_{\lambda} (H_1^T \epsilon  H_2) S + {\rm H.c.}] \ ,
\end{eqnarray}
where $\vec\sigma$ denotes the three Pauli matrices,
$g_1$, $g_2$, and $g'_1$ are the $U(1)$, $SU(2)$, and $U(1)'$ gauge coupling constants, respectively,
${\tilde Q}_1$, ${\tilde Q}_2$, and ${\tilde Q}_3$ are the $U(1)'$ hypercharges of  $H_1$, $H_2$, and $S$, respectively,
and $m_i^2$ $(i = 1, 2, 3)$ are the soft SUSY breaking masses.
In the Higgs potential, $\lambda$ and $A_{\lambda}$ may in general be complex numbers.
However, they will be assumed to be real in the subsequent discussions, as we do not consider CP violation in the Higgs sector.
The soft masses are also assumed to be real, without loss of generality, and they are eventually eliminated by imposing minimum conditions with respect to the neutral Higgs fields,
The gauge invariance of the superpotential under of $U(1)'$ requires that the three $U(1)'$ hypercharges should satisfy ${\tilde Q}_1 + {\tilde Q}_2 + {\tilde Q}_3 = 0$.

The above Higgs potential at the tree level would allow the three neutral Higgs fields $H_1^0$, $H_2^0$, and $S$ to develop the vacuum expectation values (VEVs) $v_1 (0)$, $v_2 (0)$, and $s (0)$, respectively.
Remark that these VEVs are obtained at zero temperature.
However, for simplicity, we omit the temperature dependence of these VEVs until next section where we take into account the finite temperature effect.

The tree-level Higgs potential should now be corrected by the radiative one-loop effects.
In SUSY models, the radiative corrections due to the top and stop quarks contribute most dominantly to the tree-level Higgs sector.
Besides, if $\tan\beta = v_2/v_1$ is very large, the radiative corrections due to the bottom and sbottom quarks should also be included since they become no longer negligible.
Furthermore, the radiative corrections due to the exotic quark and squark may be important since the Yukawa coupling of the exotic quark to the singlet field $S$ can be large at the electroweak scale [11].
Therefore, we take into account all the contributions from the top, bottom, exotic quark sector to the tree-level Higgs potential.

The one-loop radiative corrections are evaluated by the effective potential method [13].
We assume that the squark masses are degenerate.
Ignoring the mixings in the masses of the squarks [14], the one-loop effective potential is given by
\begin{equation}
    V_1 = \sum_{l=t,b,k} {3 {\cal M}_l^4 \over 16 \pi^2}
        \left [ {3 \over 2} + \log \left ( {{\tilde m}^2 + {\cal M}_l^2 \over {\cal M}_l^2} \right ) \right ] \ ,
\end{equation}
where $t$, $b$, and $k$, respectively are top, bottom, and exotic quark fields including the corresponding squark fields,
${\cal M}_t =  h_t |H_2|$, ${\cal M}_b = h_b |H_1|$, ${\cal M}_k = h_k |S|$ are the field-dependent quark masses,
and ${\tilde m}$ is the soft SUSY breaking mass, which is assumed that $\tilde m = 1000$ GeV $\gg m_q$ ($q $= $t$, $b$, or $k$).

The Higgs sector of the present model consists of six physical Higgs bosons: a pair of charged Higgs boson, one neutral pseudoscalar Higgs boson, and three neutral scalar Higgs bosons.
The tree-level mass of the charged Higgs boson is given by
\begin{equation}
    m_{C^{\pm}}^2 = m_W^2 - \lambda^2 v^2 + {2 \lambda A_{\lambda} s \over \sin 2 \beta} \ ,
\end{equation}
where $v = \sqrt{v^2_1 + v^2_2} = 175$ GeV and $m_W^2 = g_2^2 v^2 /2$ is the squared mass of the $W$ boson.
At the tree level, the mass of the charged Higgs boson might be either smaller or larger than the $W$ boson mass.

The tree-level mass of the neutral pseudoscalar Higgs boson is given by
\begin{equation}
    m_A^2 = {2 \lambda A_{\lambda} v \over \sin 2 \alpha} \ ,
\end{equation}
where $\tan\alpha = (v/2s) \sin 2 \beta$ implies the splitting between the electroweak symmetry breaking scale and the extra $U(1)'$ symmetry breaking scale.
Note that these tree-level masses of both the neutral pseudoscalar and the charged Higgs bosons do not receive any radiative corrections, because the squark masses are degenerate.

The tree-level squared masses of the three neutral scalar Higgs bosons are considerably affected by the radiative corrections.
Their squared masses at the one-loop level are given as the eigenvalues of the $3 \times 3$ one-loop level mass matrix, whose elements may be written as
\begin{eqnarray}
M_{11} & = & m_Z^2 \cos^2 \beta + 2 g^{'2}_1 {\tilde Q}_1^2 v^2 \cos^2 \beta
+ m_A^2 \sin^2 \beta \cos^2 \alpha + f_a(m_b^2) \cr
M_{22} & = & m_Z^2 \sin^2 \beta + 2 g^{'2}_1 {\tilde Q}_2^2 v^2 \sin^2 \beta
+ m_A^2 \cos^2 \beta \cos^2 \alpha + f_a(m_t^2) \ , \cr
M_{33} & = & 2 g^{'2}_1 {\tilde Q}_3^2 s^2 + m_A^2 \sin^2 \alpha  + f_a(m_k^2)  \ , \cr
M_{12} & = & g^{'2}_1 {\tilde Q}_1 {\tilde Q}_2 v^2 \sin 2 \beta + (\lambda^2 v^2 - m_Z^2/2) \sin 2 \beta
- m_A^2 \cos \beta \sin \beta \cos^2 \alpha \ ,  \cr
M_{13} & = & 2 g^{'2}_1 {\tilde Q}_1 {\tilde Q}_3 v s \cos \beta + 2 \lambda^2 v s \cos \beta
- m_A^2 \sin \beta \cos \alpha \sin \alpha \ , \cr
M_{23} & = & 2 g^{'2}_1 {\tilde Q}_2 {\tilde Q}_3 v s \sin \beta + 2 \lambda^2 v s \sin \beta
- m_A^2 \cos \beta \cos \alpha \sin \alpha  \ ,
\end{eqnarray}
where $m_Z^2 = (g_1^2 + g_2^2) v^2 /2$ is the squared mass of the $Z$ boson,
and the function $f_a(m_q^2)$ is defined as
\begin{equation}
    f_a(m_q^2) = {3 h_q^2 m_q^2 \over 4 \pi^2}
            \log \left ({{\tilde m}^2 + m_q^2 \over m_q^2} \right )
        + {3 \over 8 \pi^2} \left [ {4 h_q^2 m_q^4 \over {\tilde m}^2 + m_q^2 }
            - {h_q^2 m_q^6 \over ({\tilde m}^2 + m_q^2)^2} \right ] \ .
\end{equation}
We assume that the masses of three scalar Higgs bosons $S_i$ are sorted such that $m_{S_1} \le m_{S_2} \le m_{S_3}$.

\section{FINITE TEMPERATURE}

Now, let us study the temperature dependence of the Higgs potential in order to investigate the nature of the electroweak phase transition in the MSSM with an extra $U(1)'$.
We evaluate $V_T$, the temperature-dependent part of the Higgs potential at the one-loop level, using the effective potential method.
It is given as [15]
\begin{equation}
    V_T  =  \sum_{l = B, F} {n_l T^4 \over 2 \pi^2}
                \int_0^{\infty} dx \ x^2 \
                \log \left [1 \pm \exp{\left ( - \sqrt {x^2+{m_l^2(\phi_i)/T^2 }} \right ) } \right ] \ ,
\end{equation}
where $B$ and $F$ stand for bosons (${\tilde t}$, ${\tilde b}$, and ${\tilde k}$) and fermions ($t$, $b$, and $k$), and $n_t = n_b = n_k = -12$ and $n_{\tilde t} = n_{\tilde b} = n_{\tilde k} = 12$.
The negative sign is for bosons and the positive sign is for fermions.
Thus, the full Higgs potential at finite temperature at the one-loop level is given by
\begin{equation}
    V(T) = V_0 + V_1 + V_T
\end{equation}

For numerical analysis, we need to set the values of the relevant parameters of the model.
As in the previous section, the soft SUSY breaking mass is set as ${\tilde m} = 1000$ GeV.
The quark masses are set as $m_t = 175$ GeV, $m_b = 4$ GeV, and $m_k = 400$ GeV.
From these values, $m_{\tilde q} = \sqrt{{\tilde m}^2 + m_q^2}$ ($q=t,b,k$) yield the squark masses as $m_{\tilde t} = 1015$ GeV, $m_{\tilde b} = 1000$ GeV, and $m_{\tilde k} = 1077$ GeV.

Some caution should be taken for setting the values of ${\tilde Q}_i$ ($i$=1, 2, 3), the $U(1)'$ hypercharges of the Higgs doublets and the Higgs singlet.
In the MSSM with an extra $U(1)'$, the extra neutral gauge boson mass ($m_{Z'}$) and the mixing angle ($\alpha_{ZZ'}$) between the two neutral gauge bosons ($Z, Z'$) may impose strong constraints on the parameter values.
For our numerical analysis, $m_{Z'}$ is estimated to be larger than 600 GeV, and $\alpha_{ZZ'}$ smaller than $2 \times 10^{-3}$, for $\tan \beta = 3$ and $s(T=0)=500$ GeV.
Besides, as recent research has suggested [10], we impose the constraint of ${\tilde Q}_1 {\tilde Q}_2 > 0$.
Further, the $U(1)'$ gauge invariance condition requires that ${\tilde Q}_3 = -({\tilde Q}_1 + {\tilde Q}_2)$.

In this paper, we define new charges $Q_i = g_1' {\tilde Q}_i$ since ${\tilde Q}_i$ appear always together with $g_1'$.
Then, one may establish the allowed area in the ($Q_1, Q_2$)-plane by imposing the above constraints.
For $\tan \beta = 3$ and $s(T=0)=500$ GeV, the result is shown in Fig. 1,
where the small area near the point ($Q_1$, $Q_2$) = (-1, 0) and the upper right corner of Fig. 1 are the allowed areas.
The hatched region is the excluded area.
There are two specific points in Fig. 1, marked by a star ($*$) and a cross ($+$).
The values of $Q_1$ and $Q_2$ at the star-marked point correspond to the $\nu$-model of $E_6$ gauge group realizations [11].
We would take the values of $Q_1$ and $Q_2$ at the cross-marked point, namely, ($Q_1, Q_2$) = (-1, -0.1),
and hence $Q_3$ =1.1.

With these parameter values at hand, we would investigate the possibility of the strongly first-order electroweak phase transition by using two different ways.
The first method is to retain only the dominant $T^2$-proportional part from the high-temperature approximation of $V_T$, and to take account only those particles whose masses are relatively small [6].
The second method is to perform the integration in $V_T$ in numerically exact way, and to consider only the contributions of top, bottom, and exotic quarks and squarks.

\subsection{Method A}

Let us start with the high temperature approximation of $V_T$, which is expressed as [3]
\begin{eqnarray}
    V_T & \approx &\mbox{} - \sum_{i =t,b,k} n_i
    \left [{T^2 m_i^2 (\phi_i) \over 48}
    + {m_i^4 (\phi_i) \over 64 \pi^2} \log \left ({m_i^2 (\phi_i) \over c_F T^2} \right ) \right ] \cr
    & &\mbox{}
    + \sum_{i = {\tilde t}, {\tilde b}, {\tilde k}} n_i
    \left [{T^2 m_i^2 (\phi_i) \over 24} - {T m_i^3 (\phi_i) \over 12 \pi}
    - {m_i^4 (\phi_i) \over 64 \pi^2} \log \left ({m_i^2 (\phi_i) \over c_B T^2} \right ) \right ] \ ,
\end{eqnarray}
where $\log c_F = 2.64$ and $\log c_B = 5.41$.
It is known that in the SM the high temperature approximation is consistent with the exact integration of $V_T$ within 5 \% at temperature $T$ for $m_F/T < 1.6$ and $m_B/T < 2.2$, where $m_F$ and $m_B$ are respectively the fermion mass and the boson mass that participate in the potential.

We select those terms that are proportional to $T^2$ in the above expression, which become most dominant at high temperature.
Thus, we assume that the temperature at which the electroweak phase transition takes place is sufficiently high.
We also assume that the $U(1)$ and $SU(2)$ gaugino masses $M_1$ and $M_2$ in the chargino and neutralino sectors are very much larger than the other mass parameters.
We take into account the thermal effects due to the Higgs bosons, $W$, $Z$, and the extra $U(1)$ gauge boson in the boson sector, and $t$, $b$, $k$ quarks, the lighter chargino, and the three light neutralinos in the fermion sector, because their masses are relatively small as compared with temperature, similarly to the analyses of previous articles [6].
Explicitly, the $T^2$ terms in the high temperature approximation of $V_T$ can be expressed as
\begin{eqnarray}
V_T & = & {T^2 \over 24}
    \left [ 4 m_1^2 + 4 m_2^2 + 2 m_3^2 + (2 g_1^2 + 6 g_2^2 + 6 \lambda^2) (|H_1|^2 + |H_2|^2)
        + 12 \lambda^2 |S|^2 \right. \cr
    &&\mbox{}
        + 12 g^{'2}_1 ({\tilde Q}_1^2 |H_1|^2 + {\tilde Q}_2^2 |H_2|^2 + {\tilde Q}_3^2 |S|^2)
        + 2 g^{'2}_1 {\tilde Q}_1 {\tilde Q}_2 (|H_1|^2 + |H_2|^2) \cr
    & &\mbox{}
        + 2 g^{'2}_1 {\tilde Q}_2 {\tilde Q}_3 (|H_2|^2 + |S|^2)
        + 2 g^{'2}_1 {\tilde Q}_1 {\tilde Q}_3 (|H_1|^2 + |S|^2) \cr
    & &\mbox{}
        + 8 g^{'2}_1 ({\tilde Q}_1 + {\tilde Q}_2) ({\tilde Q}_1 |H_1|^2
        + {\tilde Q}_2 |H_2|^2 + {\tilde Q}_3 |S|^2  ) \cr
    & &\mbox{} \left. + 6 (h_t^2 |H_2|^2 + h_b^2 |H_1|^2 + h_k^2 |S|^2) \right ]  \ .
\end{eqnarray}

Now, the neutral scalar Higgs fields develop the temperature-dependent VEVs, $v_1 (T)$, $v_2 (T)$, and $s (T)$, which we will simply denote $v_1$, $v_2$, and $s$, respectively.
In terms of these temperature-dependent VEVs, the vacuum at finite temperature is defined as the minimum of $V(T)$ as
\begin{equation}
    \langle V(v_1,v_2,s,T) \rangle = \langle V_0 \rangle + \langle V_1 \rangle + \langle V_T \rangle \ ,
\end{equation}
where
\begin{eqnarray}
    \langle V_0 \rangle
    & = & m_1^2 v_1^2 + m_2^2 v_2^2 + m_3^2 s^2 + {g_1^2 + g_2^2 \over 8} (v_1^2 - v_2^2)^2
        + \lambda^2 (v_1^2 v_2^2 + v_1^2 s^2 + v_2^2 s^2)  \cr
    & &\mbox{}
        - 2 \lambda A_{\lambda} v_1 v_2 s + {g^{'2}_1 \over 2 }
        ({\tilde Q}_1 v_1^2 + {\tilde Q}_2 v_2^2 + {\tilde Q}_3 s^2)^2 \ , \cr
    \langle V_1 \rangle
    & = & f_b(m_t^2) + f_b(m_b^2) + f_b(m_k^2) \ , \cr
    \langle V_T \rangle
    & = & {T^2 \over 24} \left [
        4 m_1^2 + 4 m_2^2 + 2 m_3^2 + (2 g_1^2 + 6 g_2^2 + 6 \lambda^2) (v_1^2 + v_2^2)
        + 12 \lambda^2 s^2 \right. \cr
    &&\mbox{}
        + 12 g^{'2}_1 ({\tilde Q}_1^2 v_1^2 + {\tilde Q}_2^2 v_2^2 + {\tilde Q}_3^2 s^2)
        + 2 g^{'2}_1 {\tilde Q}_1 {\tilde Q}_2 (v_1^2 + v_2^2) \cr
    & &\mbox{}
        + 2 g^{'2}_1 {\tilde Q}_2 {\tilde Q}_3 (v_2^2 + s^2)
        + 2 g^{'2}_1 {\tilde Q}_1 {\tilde Q}_3 (v_1^2 + s^2) \cr
    & &\mbox{} \left.
        + 8 g^{'2}_1 ({\tilde Q}_1 + {\tilde Q}_2) ({\tilde Q}_1 v_1^2 + {\tilde Q}_2 v_2^2 + {\tilde Q}_3 s^2  )
            + 6 (h_t^2 v_2^2 + h_b^2 v_1^2 + k^2 s^2) \right ] \ .
\end{eqnarray}
In the above expressions, the function $f_b$ is defined as
\begin{equation}
    f_b(m_q^2) = {3 m_q^4 \over 16 \pi^2} \left [ {3 \over 2}
        + \log \left ({{\tilde m}^2 + m_q^2 \over m_q^2} \right ) \right ]  \ ,
\end{equation}
and the soft SUSY breaking masses at the one-loop level are given as
\begin{eqnarray}
m_1^2 & = &\mbox{} - {m_Z^2 \over 2} \cos 2 \beta - \lambda^2 (s(0)^2 + v(0)^2 \sin^2 \beta)
                  + \lambda A_{\lambda} s(0) \tan \beta   \cr
& &\mbox{} - g^{'2}_1 {\tilde Q}_1 ({\tilde Q}_1 v(0)^2 \cos^2 \beta
+ {\tilde Q}_2 v(0)^2 \sin^2 \beta + {\tilde Q}_3 s(0)^2) - f_c(m_b^2(0)) \cr
m_2^2 & = & {m_Z^2 \over 2} \cos 2 \beta - \lambda^2 (s(0)^2 + v(0)^2 \cos^2 \beta)
                  + \lambda A_{\lambda} s(0) \cot \beta   \cr
& &\mbox{} - g^{'2}_1 {\tilde Q}_2 ({\tilde Q}_1 v(0)^2 \cos^2 \beta
+ {\tilde Q}_2 v(0)^2 \sin^2 \beta + {\tilde Q}_3 s(0)^2) - f_c(m_t^2(0)) \cr
m_3^2 & = &\mbox{} - \lambda^2 v(0)^2 + {\lambda \over 2 s(0)} v(0)^2 A_{\lambda} \sin 2 \beta  \cr
& &\mbox{} - g^{'2}_1 {\tilde Q}_3 ({\tilde Q}_1 v(0)^2 \cos^2 \beta + {\tilde Q}_2 v(0)^2 \sin^2 \beta
+ {\tilde Q}_3 s(0)^2)  - f_c(m_k^2(0)) \ ,
\end{eqnarray}
where $v_1 (0)$, $v_2 (0)$, and $s(0)$ are the VEVs evaluated at zero temperature in the preceding section, $\tan \beta = v_2(0)/v_1(0)$, $v(0) = \sqrt{v_1(0)^2 + v_2(0)^2} = 175$ GeV, and the function $f_c$ is defined as
\begin{equation}
    f_c(m_q^2) =  {3 h_q^2 m_q^2 \over 16 \pi^2}
        \left [ 2 + 2 \log \left ( {{\tilde m}^2 + m_q^2 \over m_q^2}\right )
            + {m_q^2 \over {\tilde m}^2 + m_q^2} \right ] \ .
\end{equation}

Now, let us determine the critical temperature at which the electroweak phase transition takes place.
In our analysis, the critical temperature is defined by a temperature at which $\langle V (T) \rangle$ has two distinct minima with equal value, that is, a pair of degenerate vacua.
In order to have a pair of degenerate vacua, the potential $\langle V (T) \rangle$ should satisfy the minimum condition of
\begin{eqnarray}
        0 & = &
    2 m_3^2 s - 2 \lambda A_{\lambda} v_1 v_2 + 2 \lambda^2 (v_1^2 + v_2^2) s \cr
        & &\mbox{}
    + 2 g^{'2}_1 {\tilde Q}_3 s ({\tilde Q}_1 v_1^2 + {\tilde Q}_2 v_2^2 + {\tilde Q}_3 s^2)
    + 2h_k^2 m_k f_c(m_k^2) \cr
        & &\mbox{}
    +  {T^2 \over 24} s [24 \lambda^2 + 24 g^{'2}_1 {\tilde Q}_3^2
        + 20 g^{'2}_1 {\tilde Q}_3 ({\tilde Q}_1 + {\tilde Q}_2) + 12 k^2] \ ,
\end{eqnarray}
which is obtained by calculating the first derivative of the full effective potential at the finite temperature with respect to $s$.

For given parameter values at given temperature, one may solve the above minimum condition to express $s$ in terms of the other two VEVs, $v_1$ and $v_2$.
Then, by substituting $s$ into $\langle V(v_1,v_2,s,T) \rangle$, one may obtain $\langle V(v_1,v_2,T) \rangle$ which depends only on $v_1$ and $v_2$.
By inspecting the shape of $\langle V (v_1, v_2, T) \rangle$ on the ($v_1, v_2$)-plane for given parameter values at given temperature, we may determine whether it possess a pair of degenerate vacua or not.

In Fig. 2, the equipotential contours of $\langle V (v_1, v_2, T) \rangle$ are plotted on the ($v_1, v_2$)-plane, where the parameter values are set as $\tan \beta =3$, $\lambda=0.8$, $s(0)= 500$ GeV, $m_A = 1830$ GeV, and the temperature is set as $T = 100$ GeV, which is actually the critical temperature $T_c$.
One can easily spot two distinct minima of $\langle V (v_1, v_2, T) \rangle$ on the ($v_1, v_2$)-plane, namely, one at $(0, 0)$ and the other at  $(275, 640)$ GeV.
The phase of the state is symmetric at the minimum point $(0, 0)$ on the ($v_1, v_2$)-plane, whereas it is broken at $(275, 640)$ GeV.
The electroweak phase transition may take place from $(0, 0)$  to $(275, 640)$ GeV on the ($v_1, v_2$)-plane, which is evidently discontinuous and therefore it is first-order.

The distance on the ($v_1, v_2$)-plane between the two minima of $\langle V (v_1, v_2, T) \rangle$, defined as $v_c$, determines the strength of the electroweak phase transition.
The electroweak phase transition is said to be strong if  $v_c/T_c> 1$, and weak otherwise.
In Fig. 2, the distance is calculated to be
\begin{equation}
    v_c = \sqrt{(275-0)^2 + (640-0)^2}  = 696 \mbox{  \rm{(GeV)}} \ .
\end{equation}
In Fig. 2, the strength of the electroweak phase transition is about $v_c/T_c = 6.9$, which definitely tells that the electroweak phase transition is a strong one.
Therefore, the particular parameter values set for Fig. 2 yields an electroweak phase transition which is first-order as well as strong.
Note that $v_c$ does not depend on $s$, that is, we need not to know the values of $s$ at the two minima to calculate $v_c$.
Actually, $v_c$ is the VEV at the broken phase.
The masses of the neutral scalar Higgs bosons at zero temperature for the parameter values of Fig. 2
are obtained as $m_{S_1} = 56$ GeV, $m_{S_2} = 807$ GeV, and $m_{S_3} = 1827$ GeV.

We repeat the above job of analysis, varying the values of the relevant parameters.
We find that there are a large number of sets of parameter values that allow strongly first-order electroweak phase transitions.
Thus, the MSSM with an extra $U(1)'$ may accommodate the desired phase transitions for a wide region in its parameter space.
Some of the results are listed in Table 1, where $\tan \beta =3$, $s(0)= 500$ GeV, and $T = 100$ GeV are fixed as the values set in Fig. 2,
whereas $\lambda$ and $m_A$ have different values.
The set of numbers in the last row of Table 1 is the numerical result of Fig. 2.

Every set of numbers in each row of Table 1 gives $\langle V (v_1, v_2, T) \rangle$ a pair of degenerate minima, the minimum of symmetric phase at $(0, 0)$ on the ($v_1, v_2$)-plane, and the one of broken phase at a different point on the ($v_1, v_2$)-plane as given in Table 1.
The electroweak phase transition is strongly first-order.
One may easily observe in Table 1 that, as the value of $\lambda$ increases, a larger value of $m_A$ allow desired phase transitions.
On the other hand, the strength of the phase transition is reinforced if the value of $\lambda$ decreases.

The masses of the neutral scalar Higgs bosons exhibit some interesting behavior.
For a larger value of $m_A$, both $S_2$ and $S_3$ have also larger masses whereas $S_1$ has a smaller mass.
The tendency is that the strength of the phase transition is reinforced if $m_{S_1}$ increases and if $m_A$, $m_{S_2}$, and $m_{S_3}$ decrease.
In the SM, the strength of the first order electroweak phase transition decreases if its single Higgs boson mass is increased.
Also, in the MSSM, we have a weaker phase transition if the lighter one of its two scalar Higgs bosons has a larger mass.
In this regard, the tendency of our model is opposite to those of the SM or the MSSM.
One can see that this strange behavior also occurs in some parameter region of a non-minimal SUSY model, as shown in Fig. 3 of Ref. [7].

\setcounter{table}{0}
\def\tablename{}{}%
\renewcommand\thetable{TABLE 1}
\begin{table}[t]
\caption{Some sets of $\lambda$ and $m_A$ that allow strongly first-order electroweak phase transitions in the MSSM with an extra $U(1)'$, obtained by Method A.
The values of other parameters are fixed as $\tan \beta =3$, $s(0)= 500$ GeV, ${\tilde m} = 1000$ GeV, and $T_c = 100$ GeV.
The pair of numbers in the third column are the coordinates of the broken-phase minimum of $\langle V (v_1, v_2, T) \rangle$.
The coordinates of its symmetric-phase minimum is $(0,0)$ for all sets.
The three numbers in the fourth column are the masses of $S_1$, $S_2$, and $S_3$, respectively.
The number in the last column is the strength of the first-order electroweak phase transition.}
\begin{center}
\begin{tabular}{c|c|c|c|c} \hline\hline
$\lambda$ & $m_A$ (GeV) & ($v_1, v_2$) (GeV) & $m_{S_1}$, $m_{S_2}$, $m_{S_3}$ (GeV) & $v_c/T_c$  \\
\hline\hline
0.1 & 478 & (1750, 1650) & 120, 524, 792 &  26    \\
\hline
0.2 & 675 & (1400, 1500) & 118, 674, 796 &  23    \\
\hline
0.3 & 900 & (1200, 1400) & 112, 786, 908 &  18    \\
\hline
0.4 & 1109 & (870, 1200) & 104, 792, 1112 &  15    \\
\hline
0.5 & 1306 & (600, 1000) & 93, 796, 1307 &  12    \\
\hline
0.6 & 1486 & (430, 850) & 82, 800, 1485 &  8    \\
\hline
0.7 & 1660 & (340, 700) & 70, 803, 1658 &  7    \\
\hline
0.8 & 1830 & (275, 640) & 56, 807, 1827 &  6.9    \\
\hline\hline
\end{tabular}
\end{center}
\end{table}

\subsection{Method B}

The second method evaluates $V_T$ by exact integration to obtain the temperature-dependent full potential $V(T)$
at one-loop level, where the thermal effects of top, bottom,
and exotic quarks and squarks are taken into account.
The thermal effects of the gauge bosons can be a help for strengthening the first-order
electroweak phase transition, but we would omit them,
since the strength of the phase transition is already strong enough.

This method starts with the exact integral expression for $\langle V_T \rangle$ after replacing the neutral Higgs fields by their VEVs as
\begin{eqnarray}
    \langle V_T \rangle
    & = & \mbox{}
        - \sum_{l=t,b,k} {6 T^4 \over \pi^2} \int_0^{\infty} dx \ x^2 \
    \log \left [
    1 -\exp{\left ( - \sqrt {x^2+{m_l^2(v_1, v_2, s) \over T^2 }} \right )  } \right ] \cr
    & &\mbox{}
    + \sum_{l= {\tilde t}, {\tilde b}, {\tilde k}}   {6 T^4 \over \pi^2}
    \int_0^{\infty} dx \ x^2 \
    \log \left [
    1 +\exp{\left ( - \sqrt {x^2+ {{\tilde m}^2 + m_l^2(v_1, v_2, s) \over T^2 }}
    \right )  } \right ]  \ ,
\end{eqnarray}
which is different from $\langle V_T \rangle$ of Method A, while $\langle V_0 \rangle$ and $\langle V_1 \rangle$ are the same as those of Method A.
From the full $\langle V(T) \rangle = \langle V_0 \rangle + \langle V_1 \rangle + \langle V_T \rangle$, we obtain a minimum condition for degenerate vacua as
\begin{eqnarray}
        0 & = &
    2 m_3^2 s - 2 \lambda A_{\lambda} v_1 v_2 + 2 \lambda^2 (v_1^2 + v_2^2) s
        + 2 g^{'2}_1 Q_3 s ({\tilde Q}_1 v_1^2 + {\tilde Q}_2 v_2^2 + {\tilde Q}_3 s^2)   \cr
    & &\mbox{} + 2 h_k^2 m_k f_c(m_k^2) \cr
    & & \mbox{}
    + {3 T^2 \over \pi^2}
    \int_0^{\infty} dx \ x^2 \
        {2 h_k^2 s \exp (-\sqrt{x^2 + m_k^2 /T^2 })
    \over \sqrt{x^2 + m_k^2 /T^2 }
    \left[1 + \exp (-\sqrt{x^2 + m_k^2  /T^2 }) \right]}  \cr
    & &\mbox{} - {3 T^2 \over \pi^2}
    \int_0^{\infty} dx \ x^2 \ \cr
    & &\mbox{} \times
    { 2 h_k^2 s \exp (-\sqrt{x^2 + ({\tilde m}^2 + m_k^2) /T^2 })
    \over \sqrt{x^2 + ({\tilde m}^2 +  m_k^2) /T^2 }
    \left[ 1 + \exp (-\sqrt{x^2 + ({\tilde m}^2 + m_k^2) /T^2) }) \right]}   \  ,
\end{eqnarray}
where $m_k$ depends only on $s$ and is independent from $v_1$ and $v_2$.

Solving the above minimum condition is harder than solving the corresponding minimum condition of Method A.
Nevertheless, we can solve it by using the bisection method to express $s$ in terms of the other parameters.
Then, eliminating $s$ from $\langle V(T) \rangle$, we can obtain the expression for $\langle V (v_1, v_2, T) \rangle$ which depends only on $v_1$ and $v_2$.
Subsequent steps of numerical analysis are the same as the previous method.

In Fig. 3, equipotential contours of $\langle V (v_1, v_2, T) \rangle$ obtained by the present method is plotted
on the ($v_1, v_2$)-plane, where the parameter values are set slightly different from the previous method:
$\tan \beta =3$, $\lambda=0.8$, $s(0)= 500$ GeV, $m_A = 1780$ GeV, and $T = 100$ GeV.
The shape of the equipotential contours of Fig. 3 is almost the same as that of Fig. 2.
One can see that there are two distinct minima in Fig. 3, just like Fig. 2: one at $(0,0)$, and the other at $(165,440)$ GeV on the  ($v_1, v_2$)-plane, indicating that the phase transition is first order.
The strength of the first-order phase transition is strong, since $v_c/T_c = 4.7$.
The masses of the three scalar Higgs bosons are evaluated at zero temperature
as $m_{S_1} = 82$ GeV, $m_{S_2} = 804$ GeV, and $m_{S_3} = 1777$ GeV.

Comparing Fig. 3 with Fig. 2, one may safely remark that Method A and Method B lead qualitatively the same results.
Either method, whether $\langle V_T \rangle$ is calculated by direct integration or is simplified by high-temperature approximation,
and whether the participating particles at the one-loop level are somewhat exhaustive or selective,
we find that the MSSM with and extra $U(1)'$ allows strongly first-order electroweak phase transitions for certain region in its parameter space.

We repeat the numerical analysis by varying the parameter values. and some of the results are listed in Table 2.
Like in Table 1, $\tan \beta =3$, $s(0)= 500$ GeV, and $T = 100$ GeV are fixed, whereas $\lambda$ and $m_A$ are varied.
The set of numbers in the last row of Table 2 is the numerical result of Fig. 3.
Comparing Table 2 with Table 1, one may notice that the numbers are slightly different from each other
but the general behavior of the two tables is exactly the same.

\setcounter{table}{0}
\def\tablename{}{}%
\renewcommand\thetable{TABLE 2}
\begin{table}[t]
\caption{Some sets of $\lambda$ and $m_A$ that allow strongly first-order electroweak phase transitions in the MSSM with an extra $U(1)'$, obtained by Method B. Other descriptions are the same as Table 1.}
\begin{center}
\begin{tabular}{c|c|c|c|c} \hline\hline
$\lambda$ & $m_A$ GeV & ($v_{1B}, v_{2B}$) GeV & $m_{S_i}$ GeV & $v_c/T_c$  \\
\hline\hline
0.1 & 462 & (1600, 1600) & 121, 468, 791 &  22    \\
\hline
0.2 & 663 & (1400, 1400) & 118, 662, 795 &  19    \\
\hline
0.3 & 885 & (1100, 1100) & 113, 785, 894 &  15    \\
\hline
0.4 & 1095 & (800, 1200) & 106, 792, 1098 &  14    \\
\hline
0.5 & 1287 & (680, 990) & 97, 796, 1288 &  12    \\
\hline
0.6 & 1457 & (400, 750) & 91, 799, 1456 &  8    \\
\hline
0.7 & 1620 & (300, 600) & 86, 801, 1618 &  6    \\
\hline
0.8 & 1780 & (165, 440) & 82, 804, 1777 &  4.7    \\
\hline\hline
\end{tabular}
\end{center}
\end{table}

\section{DISCUSSIONS AND CONCLUSIONS}

We investigate the MSSM with an extra $U(1)'$ if it could accommodate strongly first-order
electroweak phase transitions to provide sufficient baryon asymmetry,
for reasonable masses of scalar Higgs bosons.
To do so, we need the temperature-dependent part of the Higgs potential at the one-loop level.
Explicitly, its expression is obtained by two complementary methods:
Method A employs high-temperature approximation and retains only the most dominant $T^2$ terms,
and takes into account the thermal effects at the one-loop level of various participating particles.
On the other hand, method B performs numerical integrations, and the thermal effects of top, bottom,
and exotic quarks and squarks are accounted for.

Both methods lead us to essentially the same conclusion:
the strongly first-order electroweak phase transition is possible in the MSSM with an extra $U(1)'$,
for a wide region in its parameter space.
The masses of the scalar Higgs bosons are obtained within reasonably acceptable ranges.
Accordingly, we may expect that the MSSM with an extra $U(1)'$ can explain the baryon asymmetry of the universe.

We remark that the MSSM with an extra $U(1)'$ exhibits an interesting behavior with respect to the correlation between the strength of the phase transition and the Higgs boson masses.
The MSSM with an extra $U(1)'$ is opposite to the SM or to the MSSM in the sense that the mass of the lightest scalar Higgs boson increases when the strength of the strongly first-order electroweak phase transition becomes stronger.
In the SM, its single Higgs boson has a larger mass when the strength of the first order electroweak phase transition decreases.
In the MSSM, we also have a larger mass for the lighter one of its two scalar Higgs bosons when the phase transition becomes weaker.

\vskip 0.3 in
\noindent
{\large {\bf ACKNOWLEDGMENTS}}
\vskip 0.2 in
This research is supported by KOSEF through CHEP.
The authors would like to acknowledge the support from KISTI (Korea Institute of Science and Technology Information) under "The Strategic Supercomputing Support Program" with Dr. Kihyeon Cho as the technical supporter.
The use of the computing system of the Supercomputing Center is also greatly appreciated.


\vfil\eject


{\large {\bf FIGURE CAPTION}}

\vskip 0.2 in
\noindent
FIG. 1. :
The allowed area in the ($Q_1, Q_2$)-plane.
For $\tan \beta = 3$ and $s(T=0)=500$ GeV, the small area near the point ($Q_1, Q_2$) = (-1, 0)
and the upper right corner are the allowed areas, whereas the hatched region is the excluded area.
There are two specific points, marked by a star ($*$) and a cross ($+$).
The values of $Q_1$ and $Q_2$ at the star-marked point correspond to the $\nu$-model of $E_6$ gauge group realizations.
The values of $Q_1$ and $Q_2$ at the cross-marked point are
($Q_1, Q_2$) = (-1, -0.1), and hence $Q_3$ =1.1. In our discussions, we choose this point.

\vskip 0.2 in
\noindent
FIG. 2. : The plot of the equipotential contours of $\langle V (v_1, v_2, T) \rangle$ on the ($v_1, v_2$)-plane, obtained by Method A.
The parameter values are set as $\tan \beta =3$, $\lambda=0.8$, $s(0)= 500$ GeV, $m_A = 1830$ GeV, and the temperature is set as $T = 100$ GeV, which is actually the critical temperature $T_c$.
Notice two distinct minima of $\langle V (v_1, v_2, T) \rangle$ on the ($v_1, v_2$)-plane: $(0, 0)$ where the phase of the state is symmetric,
and $(275, 640)$ GeV, where the phase of the state is broken.
The electroweak phase transition may take place from $(0, 0)$  to $(275, 640)$ GeV on the ($v_1, v_2$)-plane,
which is evidently discontinuous and therefore it is first order.
The distance between the two minima is $v_c = 696$ GeV, indicating that the strength of the first-order phase transition is strong ($v_c/T_c > 1$).
The masses of the three scalar Higgs bosons are obtained as $m_{S_1} = 56$ GeV, $m_{S_2} = 807$ GeV,
and $m_{S_3} = 1827$ GeV.

\vskip 0.2 in
\noindent
FIG. 3. : The plot of the equipotential contours of $\langle V (v_1, v_2, T) \rangle$ on ($v_1, v_2$)-plane, obtained by Method B.
The parameter values are set as $\tan \beta =3$, $\lambda=0.8$, $s(0)= 500$ GeV, $m_A = 1780$ GeV, and $T_c = 100$ GeV.
The coordinates of two minima are: $(0,0)$ and $(165,440)$ GeV.
The distance between the two minima is $v_c = 470$ GeV, thus the electroweak phase transition between the two minima is strongly first-order.
The masses of the three scalar Higgs bosons are obtained as $m_{S_1} = 82$ GeV, $m_{S_2} = 804$ GeV,
and $m_{S_3} = 1777$ GeV.

\vfil\eject

\setcounter{figure}{0}
\def\figurename{}{}%
\renewcommand\thefigure{FIG. 1}
\begin{figure}[t]
\begin{center}
\includegraphics[scale=0.6]{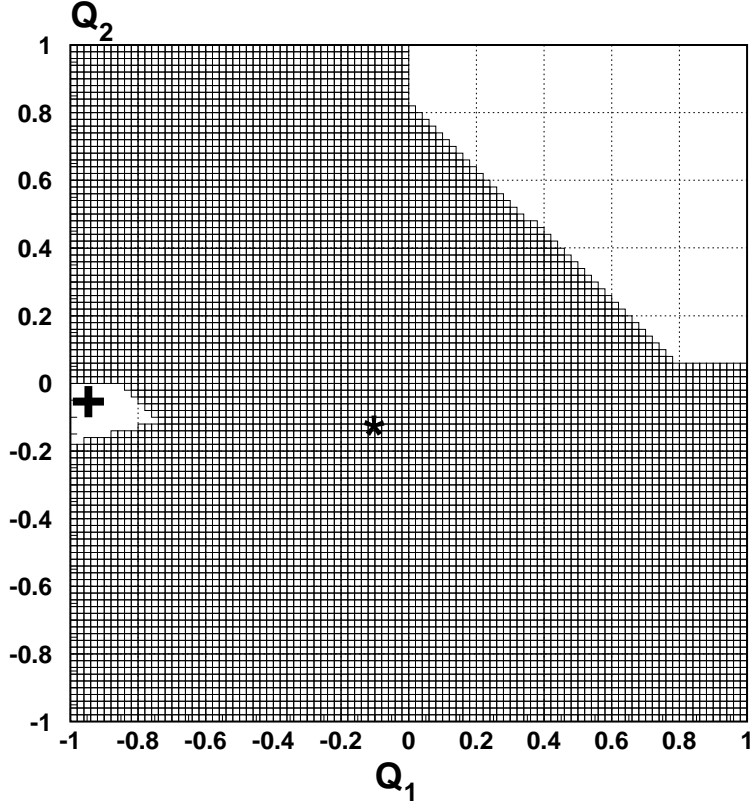}
\caption[plot]{The allowed area in the ($Q_1, Q_2$)-plane.
For $\tan \beta = 3$ and $s(T=0)=500$ GeV, the small area near the point ($Q_1, Q_2$) = (-1, 0)
and the upper right corner are the allowed areas, whereas the hatched region is the excluded area.
There are two specific points, marked by a star ($*$) and a cross ($+$).
The values of $Q_1$ and $Q_2$ at the star-marked point correspond to the $\nu$-model of $E_6$ gauge group realizations.
The values of $Q_1$ and $Q_2$ at the cross-marked point are
($Q_1, Q_2$) = (-1, -0.1), and hence $Q_3$ =1.1.
In our discussions, we choose this point.}
\end{center}
\end{figure}

\setcounter{figure}{0}
\def\figurename{}{}%
\renewcommand\thefigure{FIG. 2}
\begin{figure}[t]
\begin{center}
\includegraphics[scale=0.6]{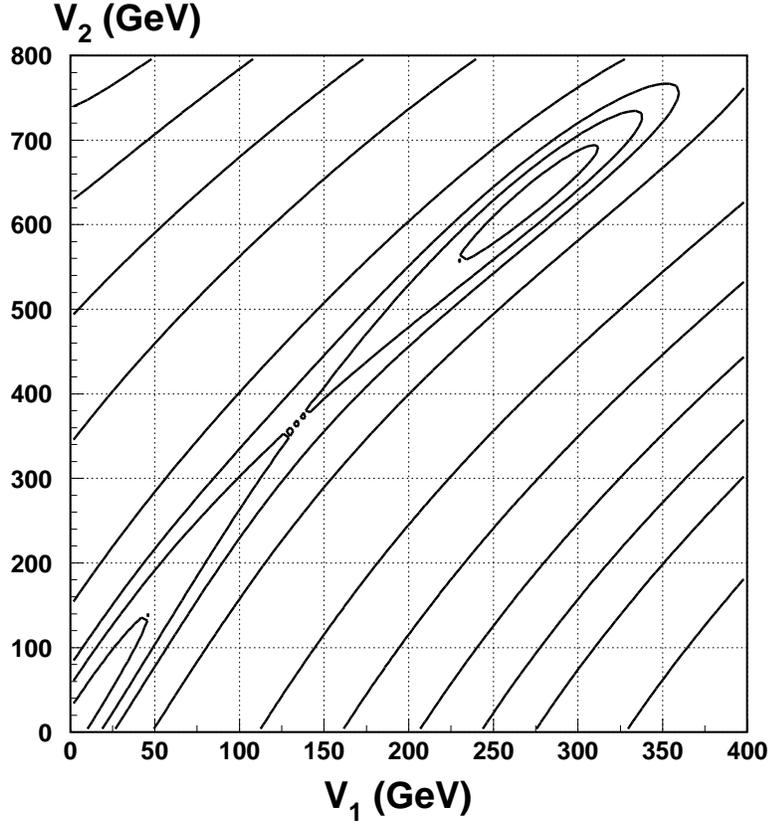}
\caption[plot]{The plot of the equipotential contours of $\langle V (v_1, v_2, T) \rangle$ on the ($v_1, v_2$)-plane, obtained by Method A.
The parameter values are set as $\tan \beta =3$, $\lambda=0.8$, $s(0)= 500$ GeV, $m_A = 1830$ GeV, and the temperature is set as $T = 100$ GeV, which is actually the critical temperature $T_c$.
Notice two distinct minima of $\langle V (v_1, v_2, T) \rangle$ on the ($v_1, v_2$)-plane: $(0, 0)$ where the phase of the state is symmetric, and $(275, 640)$ GeV, where the phase of the state is broken.
The electroweak phase transition may take place from $(0, 0)$  to $(275, 640)$ GeV on the ($v_1, v_2$)-plane, which is evidently discontinuous and therefore it is first order.
The distance between the two minima is $v_c = 696$ GeV, indicating that the strength of the first-order phase transition is strong ($v_c/T_c > 1$).
The masses of the three scalar Higgs bosons are obtained as $m_{S_1} = 56$ GeV, $m_{S_2} = 807$ GeV,
and $m_{S_3} = 1827$ GeV.}
\end{center}
\end{figure}

\setcounter{figure}{0}
\def\figurename{}{}%
\renewcommand\thefigure{FIG. 3}
\begin{figure}[t]
\begin{center}
\includegraphics[scale=0.6]{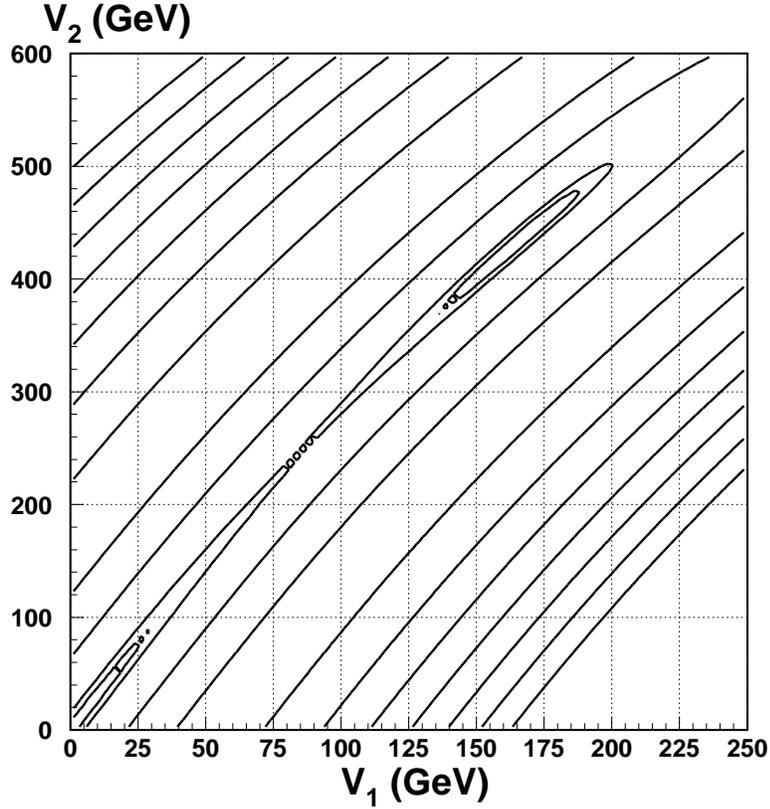}
\caption[plot]{The plot of the equipotential contours of $\langle V (v_1, v_2, T) \rangle$ on ($v_1, v_2$)-plane, obtained by Method B.
The parameter values are set as $\tan \beta =3$, $\lambda=0.8$, $s(0)= 500$ GeV, $m_A = 1780$ GeV, and $T_c = 100$ GeV.
The coordinates of two minima are: $(0,0)$ and $(165,440)$ GeV.
The distance between the two minima is $v_c = 470$ GeV, thus the electroweak phase transition between the two minima is strongly first-order.
The masses of the three scalar Higgs bosons are obtained as $m_{S_1} = 82$ GeV, $m_{S_2} = 804$ GeV,
and $m_{S_3} = 1777$ GeV.}
\end{center}
\end{figure}

\end{document}